\documentclass[reprint,amsmath,amssymb,aps]{revtex4-1}
\usepackage{graphicx}% Include figure files
\usepackage{dcolumn}% Align table columns on decimal point
\usepackage{bm}% bold math
\usepackage{amsfonts}
\usepackage{xcolor}
\begin{document}
\preprint{APS/123-QED}
\title{Granular flow from silos with rotating orifice}% Force line breaks with \\
\author{Kiwing To}
\email{ericto@gate.sinica.edu.tw}
\author{Yi-Kai Mo}
\author{Jung-Ren Huang}
\affiliation{Institute of Physics, Academia Sinica, Taipei, Taiwan R.O.C. }

\author{Yun Yen}
\affiliation{Department of Physics, National Taiwan University, Taipei, Taiwan R.O.C.}
\date{\today}% It is always \today, today,

\begin{abstract}
For granular materials falling through a circular exit at the bottom of a silo, no continuous flow can be sustained when the diameter $D$ of the exit is less than 5 times the characteristic size of the grains. 
If the bottom of the silo rotates horizontally with respect to the wall of the silo, finite flow rate can be sustained even at small $D$. 
We investigate the effect of bottom rotation to the flow rate of a cylindrical silo filled with mono-disperse plastic beads of $d=6$ mm diameter. 
We find that the flow rate $W$ follows Beverloo Law down to $D=1.2d$ and that $W$ increases with the rotation rate $\omega$ in the small exit regime.
If the exit is at an off-center distance $R$ from the axis of the silo, $W$ increases with rate of area swept by the exit. 
On the other hand, when the exit diameter is large, $W$ decreases with rotation speed at small $\omega$ but increases with $\omega$ at large $\omega$. 
Such non-monotonic behavior of $W$ on rotation speed may be explained as a gradual change from funnel flow to mass flow due to the shear at the bottom of the silo.
%Such non-monotonic behavior of $W$ on $\omega$ is found to be associated with change in the flow pattern (from funnel flow to mass flow) in the silo due to shear at the bottom.
\end{abstract}

\maketitle

Discharge of granular materials from hoppers and silos is a process which has been studied for a long time.
Obviously, the outflow rate of grains from silos and hoppers increases with the size of orifice.  
%The flow of granular materials through these appliances is controlled (or limited) by the outlet. 
%Clearly the outlet size is the most important factor that determines the outflow rate. 
%The relationship between the flow rate and the orifice size has been known for more than half century. 
The empirical observation between flow rate $W$ and orifice size $D$: $W\propto (D-D_o)^{5/2}$ by Beverloo \cite{Beverloo61} has been confirmed when $D$ is much larger than the characteristic size of the grains. 
Other factors, such as friction, elasticity and shape of the grains \cite{Ashour17,Hong17,Tang16}, hopper geometry \cite{Nedderman82, Lopez19}, external perturbations \cite{To02,Corwin08}, ..., etc., have, in general, only minor effect on flow rate except in small orifice sizes.
In principle, Beverloo law breaks down for small exit size when the flow is clogged stochastically \cite{To01,Zuriguel03}. 
The experimentally observed clogging transition probability can be fitted equally well to either an exponential decaying function or an algebraic function \cite{Zuriguel05, To05, Janda08, Thomas15}.
%if the exit size is larger than a critical value and 0 otherwise \cite{Zuriguel05, To05, Janda08, Thomas15}. 
These results led to a fundamental question of the existence of the no-clog regime when exit size is larger than a critical value. 
In contrast, unclogging transition may involve collective rearrangement of the grains \cite{Merrigan18} in the packing with a much larger length scale than exit size. 
Despite a large amount of experimental, theoretical and computational researches in clogging and unclogging in granular flow through bottle neck, quantitative theories that successfully explain these two stochastic transitions are yet to be found.

Since hoppers and silos are very common industrial and agricultural appliances that are used as transporting or distributing granular materials, there are devices (vibrator, air cannons, … etc) invented to get rid of clogging and keep the material flowing continuously. 
Recently, it was found that adding small motion of the exit is an effective way to avoid clogging in the small exit regime \cite{To17}. 
However, the effect of exit motion to flow rate is not clearly understood yet.
It has been known that clogging is due to formation of an arch in two-dimensional silos (or a dome in three-dimensional silos) at the exit that blocks the flow \cite{To01,Zuriguel03,Ashour17}. 
Hence, to prevent clogging it is crucial to prevent formation of the arch and/or break the arch once it is formed.
As demonstrated by To and Tai \cite{To17}, an efficient way is to let go the bases of the arch that blocks the flow in a two-dimensional silo by moving the exit beneath them.
%One of the most efficient ways is to let go the bases of the arch that block the flow in a two-dimensional silo by moving the exit beneath them, as demonstrated in Ref. \cite{To17}. 
% in two-dimensional silos with oscillating exit. 
However, the motion of the exit not only destroys the arch by going under the bases of the arch but also shears the materials in the two-dimensional silo. 

In order to study the effect of shear at the bottom to granular flow through a silo, we constructed a three-dimensional silo with a rotatable bottom. 
When the circular exit is at the center of the bottom of the silo, rotating the bottom induces shear to the material in the silo while the exit remains stationary.
We find that at small exit sizes, rotation of the bottom increases the flow rate as expected from simple physical argument and observed by numerical simulations \cite{Corwin08,hilton2010effect}.
Surprisingly, when the exit size is large, rotation of the bottom reduces the flow rate at small rotation speed but increases the flow rate at large rotation speed.
We observe that the non-monotonic dependence of flow rate on rotation speed may be related to change in the flow pattern---from funnel flow to mass flow in the silo. 

In the rest of the paper, we shall give an account of the experimental setup and procedures. 
Then the measurement of flow rate at different sizes, positions and rotation speeds of the exit will be presented along with our understanding of the observed behavior.
Afterward, we shall discuss possible physical pictures that may lead to the observed minimum in flow rate due to rotation of the exit of the silo.

\begin{figure}[tbph] \centering \includegraphics[width=8.5cm,height=6.5cm]{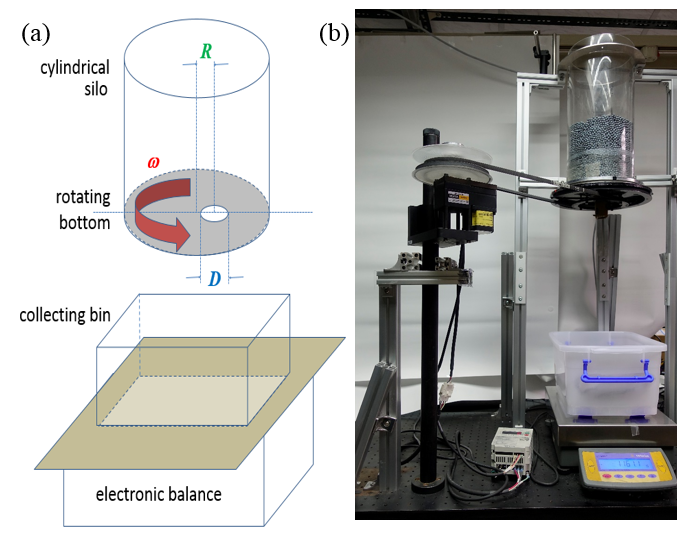}
\caption{(a) Schematic diagram of the experimental setup. (b) Photograph of the actual Experiment setup. }
\label{fig:ExptSetup}
\end{figure}

Fig. \ref{fig:ExptSetup}(a) is a schematic diagram of the experimental setup which consists of a transparent acrylic cylindrical silo of 19 cm diameter securely mounted on an aluminum frame.
The bottom of the silo can rotate about the axis of the silo with respect to the cylindrical wall. 
The rotation speed can be controlled from 0.08 revolution per second (rps) to 1.0 rps by a DC motor (Oriental Motor BLFD30A2) through a belt pulley mechanism.
The exit to the silo is a circular orifice of diameter $D$ at a distance $R$ from the center of the silo.
At the beginning of an experiment, 8000 g (about $3\times 10^4$) of plastic beads (diameter 0.58 mm and mass 0.27 g) are loaded to the silo with the exit blocked.
The bottom of the silo is set to rotate at the designated speed $\omega$ before the exit is unblocked.
The beads from the silo fall through the exit into the collecting bin whose weight is monitored every 0.1 s by an electronic balance (Satorius CP-34001P) which is configured to output only the mass of the beads in the bin to a personal computer (Asus Eee BOX B202).
%which is configured to send only the weight of the beads in the bin to a personal computer. 
Fig. \ref{fig:ExptSetup}(b) shows the time variation of the reading $m$ from the electronic balance for four different sets of the three control parameters: exit diameter $D$, off-center distance $R$ and rotation speed $\omega$. 
One can see that the total mass of the beads discharged from the silo increases linearly with time until the silo is almost empty.
From these data, it is easy to find that $W$ increases with increasing $D$, $R$ as well as $\omega$.
A quantitative measurement of $W$ is obtained by linear regression using the data in the early stage of discharge when $m <$ 2000 g. 

\begin{figure}[tbph] \centering \includegraphics[width=8.5cm,height=6.5cm]{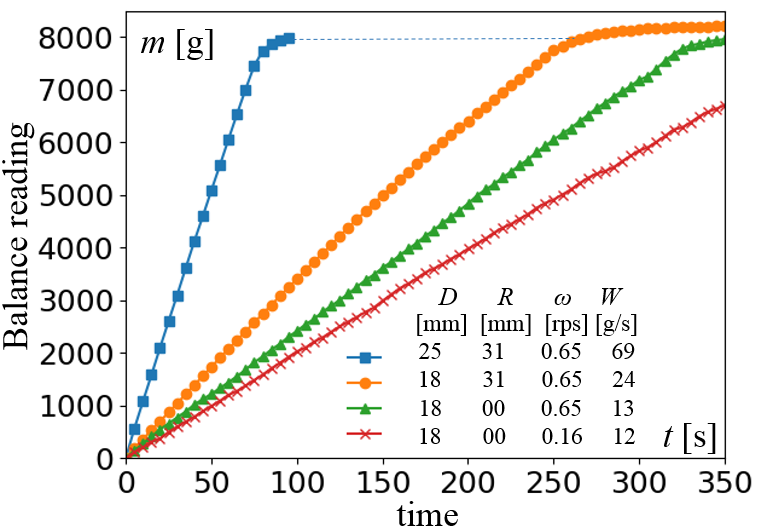}
\caption{Time records of the reading $m$ from the electronic balance during discharge at different control parameters ( $D$, $R$ and $\omega$) and the flow rates $W$ measured from $m$.}
\label{fig:Balance reading}
\end{figure}

\begin{figure}[tbph] \centering \includegraphics[width=8.5cm,height=6.5cm]{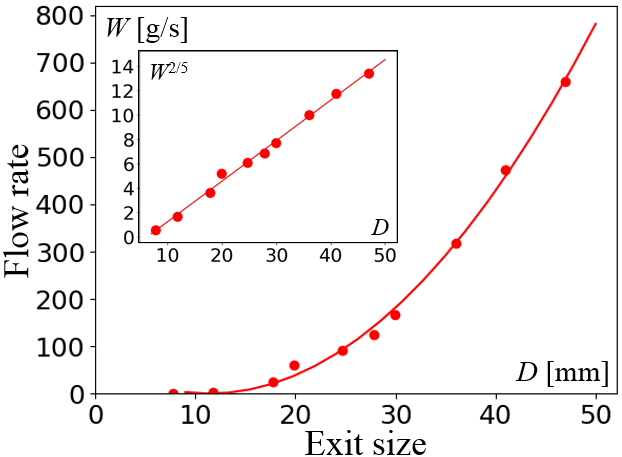}
\caption{Flow rate $W$ versus exit diameter $D$ for $R=0$ mm and $\omega= 1/6$ rps. Inset: same data plotted in $W^{2/5}$ versus $D$. Lines in this graph and the inset are the fitted curves of the data to Beverloo law $W=0.52(D-11.4)^{5/2}$.}
\label{fig:beverloo law}
\end{figure}

%\section{\label{results}Results and Discussions}
\begin{figure}[tbph]
\includegraphics[width=8.5cm,height=6.5cm]{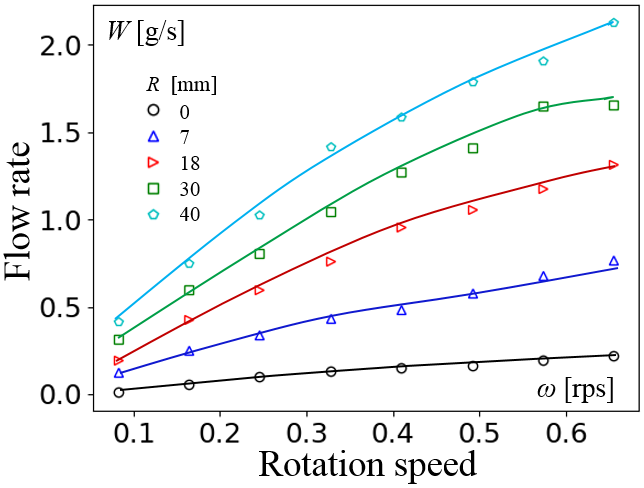}
    \caption{Graph of flow rate $W$ versus rotation speed $\omega$ at different off-center distance $R$ for silo with exit diameter $D$ = 7.8 mm. Lines in the graph are guide to the eyes only.}
    \label{fig:small exit}
\end{figure}

Fig. \ref{fig:ExptSetup}(c) shows the variation of flow rate $W$ with respect to exit diameter $D$ when the exit is at the center of the silo (i.e., $R=0$) and the rotation speed set at $\omega=1/6$ rps.
The nonlinear growth of $W$ with $D$ can be fitted to the Beverloo Law $W\propto (D-D_o)^{5/2}$ as illustrated by the inset in which the data fall on a straight line when $W^{2/5}$ is plotted against $D$.
Hence Beverloo Law is valid for silo with rotating bottom. 
It should be pointed out that, if the bottom does not rotate (i.e., $\omega=0$), continuous flow cannot sustain (i.e., $W$ vanishes) for $D\lessapprox 30$ mm. 
This implies that bottom rotation extends the validity of the Beverloo Law down to $D$ = 7.8 mm which is only 1.3 time the diameter of the bead.
This finding is consistent to those obtained by Thomas and Durian \cite{Thomas13} and Mankoc $et$. $al$ \cite{Mankoc07, Mankoc09}. 
In their works, the silo exit did not move and the flow rate in the clogging regime was taken as the mean value of the ratio of the weight of the grains discharged before clogging divided by the duration of flow in repeated flow and clog events.  

\begin{figure}[tbph]
\includegraphics[width=8.5cm,height=6.5cm]{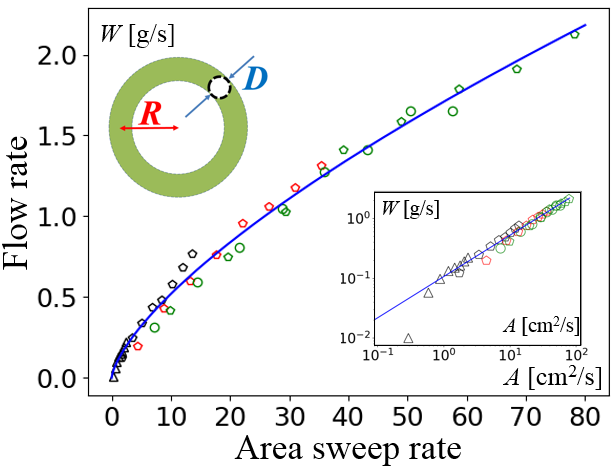}
\caption{Flow rate $W$ versus area sweep rate $A$ for exit diameter $D$ = 7.8 mm. The line through the data is a fitted curve: $W=0.1A^{0.69}$. Inset: log-log plot of $W$ versus $A$. The top-left schematic figure shows the region swept by the exit in one complete revolution when $R > D/2$.}
\label{fig:sweep area}\end{figure}

To investigate the effect of bottom rotation and exit position to flow rate, we measure the flow rate $W$ when the center of the exit is set to a distance $R$ from the center of the silo.
The results for exit diameter of $D$ = 7.8 mm are shown in Fig. \ref{fig:small exit}.
While $W$ increases with $\omega$ as expected, the enhancement of flow rate by exit rotation is more effective when the exit is farther away from the center of the silo.
If we compare the flow rate at $R$ = 40 mm to that at the center (i.e., $R$ = 0 mm), an increase by a factor of 10 in $W$ is found. 
Although $W$ increases with $R$ and $\omega$ in a nontrivial way, it may be understood by a simple arch breaking mechanism similar to that proposed in Ref. \cite{To17}.
When the exit is small, arches that block the flow will form readily after the previous arch is broken when the exit passes under the bases of the arch.
For each arch breaking event, a certain number of beads, which should be proportional to the area of the exit, will be discharged. 
Since arch breaking rate should increase with the rotation speed of the exit, it is reasonable that the flow rate should also increase with the area swept by the exit per unit time.
The region swept by the exit in one complete revolution is an annulus (see the schematic diagram in Fig. \ref{fig:sweep area}) of area $\pi(R^2-(D/2)^2)$ if $R > D/2$, or a circle of area $\pi(R+D/2)^2$ if $R \leq D/2$. 
Hence the area sweep rate of the exit is $A=\pi(R^2-(D/2)^2)\omega$ for $R > D/2$ or $\pi(R+D/2)^2\omega$ otherwise.
Thus the flow rate should depend only on the area sweep rate so that $W$ should collapse on a single curve when plotted against $A$. 

\begin{figure}[tbph]
\includegraphics[width=8.5cm,height=6.5cm]{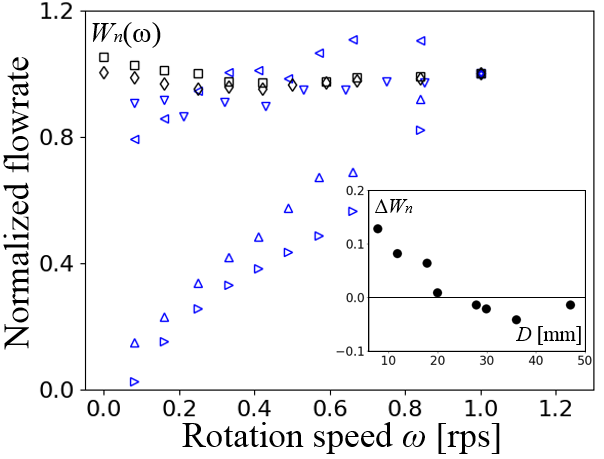}
    \caption{Normalized flow rate $W_n(\omega)$ versus rotation speed $\omega$ for exit diameter $D$ = 7.8 mm ($\color{blue}{\triangleright}$), 11.8 mm  ($\color{blue}{\triangle}$), 17.8 mm ($\color{blue}{\triangleleft}$), 19.9 mm ($\color{blue}{\triangledown}$), 29.9 mm ($\lozenge$) and 36.0 mm ($\square$). 
    Inset: Normalized flow rate difference $\Delta W_n$ for different exit diameters.
    %The definition of normalized flow rate difference is $\Delta W_n=\delta W/W_{\omega=1Hz}$, and $\delta W$ is the flow rate difference of 0.16(rps) and 0.08(rad/sec). 
    %$W_{\omega=1Hz}$ is defined from extended Beverloo law which the experiment conditions are $\omega \approx 1rps$ and $R$ = 0.
    }
    \label{fig:slope}
\end{figure}

When the data in Fig. \ref{fig:small exit} are plotted against the area sweep rate, they indeed fall on a single curve which can be fitted to a power law with an exponent of 0.69 (see Fig. \ref{fig:sweep area}). 
Flow rate data for other exit sizes behave similarly with the fitted exponent decreasing with increasing exit diameter.
Since the exponent, even though the fitting of $W$ to a power law is empirical, indicates how strongly $W$ depends on rotation speed, the influence of rotation speed to flow rate decreases with the diameter of the exit.
%In other words, the influence of rotation speed to flow rate decreases with the diameter of the exit.
%This trend can be seen by plotting the normalized flow rate $W/W_0$, with $W_0$ being the flow rate at $\omega = 2$ rad/s, in the inset of Fig. \ref{fig:slope}.
%This trend can be seen by plotting the normalized flow rate $W_n(\omega)\equiv W(\omega)/W(\omega=$2 rad/s) in the inset of Fig. \ref{fig:slope}.
This trend can be seen by plotting the normalized flow rate $W_n(\omega)\equiv W(\omega)/W(1)$ in Fig. \ref{fig:slope}.
Here $W(\omega)$ is used to emphasize the explicit dependence of flow rate $W$ on rotation speed $\omega$.
In this graph the flow rate data fall into two groups: small exit ($D<24$ mm) of strong dependence on $\omega$ and large exit ($D>25$ mm) of weak dependence on $\omega$.
To compare quantitatively how flow rate is affected by exit rotation, we calculate the normalized flow rate difference $\Delta W_n\equiv W_n(0.16) - W_n(0.08)$ which is an approximation to the rate of change in flow rate with respect to rotation speed at small $\omega$. 
%at $\omega=$ 1 rad/s and 0.5 rad/s.
%We calculate the differential normalized flow rate, defined $S\equiv W/W_0 $, as an indicator of the effect of rotation.
We find that $\Delta W_n$ decreases with exit size and turns negative when $D>25$ mm.
Hence, exit rotation enhances flow rate for small exit sizes but reduces flow rate when exit size is larger than a critical value $D_c\approx$ 25 mm.

\begin{figure}[tbph] \centering
\includegraphics[width=8.5cm,height=6.5cm]{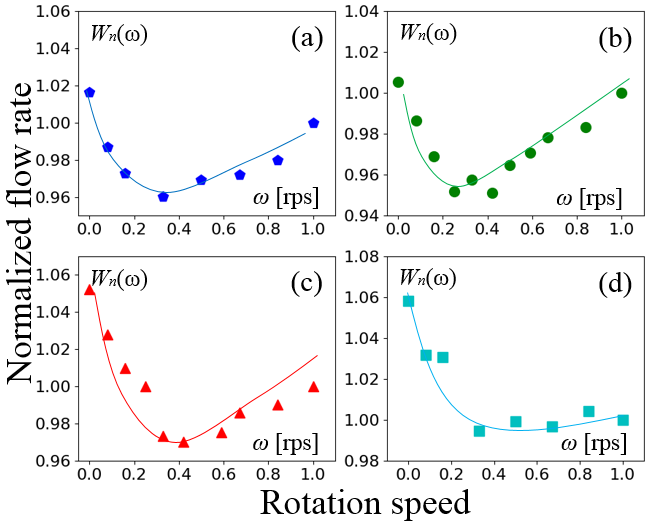}
    \caption{Normalized flow rate $W_n(\omega)$ versus rotation speed $\omega$ for silo with exit diameter $D$ = 27.8 mm (a), 29.9 mm (b), 36.0 mm (c), and 47.0 mm (d). The lines in the graphs are guide to the eyes only.}
    \label{fig:medium exit}
\end{figure}

While the flow rate enhancement at small exit sizes has been explained in previous paragraphs, the physics underlying the reduction of flow rate at large exit sizes is not clear.
The explanation given in Ref. \cite{To17} for two-dimensional silo with oscillating exit cannot apply to our three-dimensional silo with rotating exit at the center of the silo because the rotation of the exit does not involve motion of the edge of the exit in the radial direction. 
A closer look at the flow rate for large exit sizes leads to the discovery of a minimum in $W_n(\omega)$ at $\omega = \omega_c\approx 0.4$ rps, as illustrated in Fig. \ref{fig:medium exit}.

\begin{figure}[tbph]
\includegraphics[width=8.5cm,height=6.5cm]{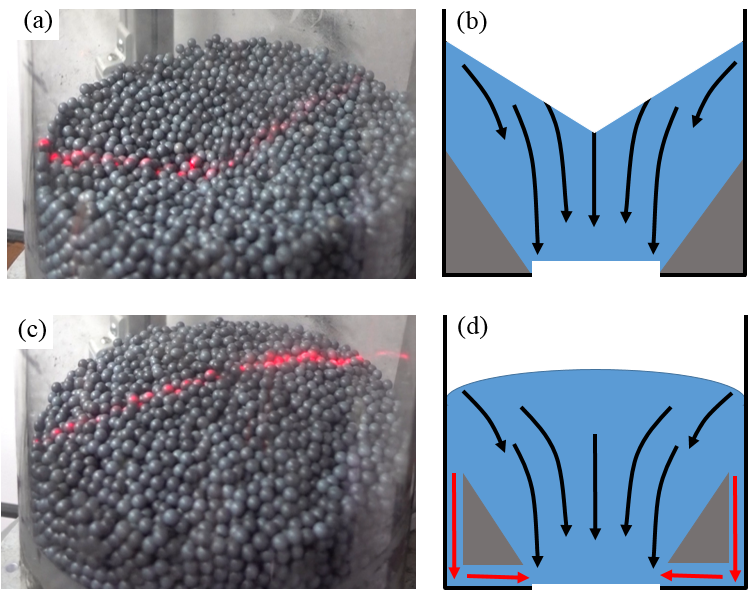}
    \caption{The Shape of the top layer of the bead packing in the discharging silo with exit diameter $D$ = 41 mm and the schematic diagram of the possible flow patterns inside the silo for ((a), (b)) $R$ = 0 mm, $\omega$ = 0 rps and ((c), (d)) $R$ = 0 mm, $\omega$ = 1 rps.}
    \label{fig:flow pattern}
\end{figure}

The presence of a minimum in $W_n(\omega)$ implies qualitative change in the discharge process from flow reduction for small rotation speed to flow enhancement for large rotation speed.
Such change seems to be related to a cross-over from funnel flow to mass flow as observed from the shape of the top surface of the grains in the silo.
When the exit is stationary, we observe a depression at the center of the top surface as shown by a light sheet generated from the beam of laser pointer through a glass rod (see Fig. \ref{fig:flow pattern}(a)).
This is a typical feature of funnel flow \cite{Pacheco17} such that an active flow (blue) region forms above the exit with a stagnant zone at the periphery as shown schematically by the grey region in Fig. \ref{fig:flow pattern}(b). 
On the other hand, When the bottom is rotating, the top surface is flat and depression is observed at the wall of the silo as shown in Fig. \ref{fig:flow pattern}(c). 
In fact, beads at the wall of the silo near the bottom are not stationary but move downward and then move inward when they reach the bottom. 
Hence a current along the wall and the bottom of the silo exists during the discharge as shown schematically in Fig. \ref{fig:flow pattern}(d)). 
When this current (red) arrives at the exit, its direction is perpendicular to that of the current (black) along the central part of the silo. 
Collisions between the beads in these two currents will generate upward impulses that reduce the overall flow rate similar to that observed in two-dimensional silo with oscillation exit \cite{To17}.
As the rotation speed increases, the bottom current increases and the out flow rate reduction increases.
On the other hand, stagnation region should shrink with increasing rotation speed.
Presumably, stagnation region vanishes before rotation speed reaches $\omega_c$, when the cross-over from funnel flow to mass flow is complete.
Further increase in $\omega$ leads to higher degree of fluidization and higher flow rate. 

%\section{\label{summary}Summary and Conclusions}
To summarize, we report experimental results on granular flow through a circular orifice in a silo with a rotatable bottom.
We find that exit rotation prevents permanent clogging at small exit diameters and the Beverloo law, which relates the flow rate $W$ to the exit diameter $D$, is found to be valid for $D$ down to 1.3 times the diameter of the grains.
In a silo with rotating bottom and orifice at a distance from the center of the silo, flow rate increases, not only with the exit diameter but also with the area swept by the orifice per unit time. 
Surprisingly, the flow rate $W$ goes through a minimum when the rotation speed $\omega$ increases. 
%$W(\omega)$ for large exit diameter decreases with rotation speed $\omega$ at small $\omega$ but increases with $\omega$ at large $\omega$. 
The non-monotonic variation of $W$ with $\omega$ can be explained by a change from funnel flow to mass flow as evident from the shape of the top surface of the grain packing in the silo.
Preliminary results \cite{Pongo,Hidalgo} from numerical simulations using discrete element method agree qualitatively to the change in flow pattern due to bottom rotation. 
Details of the numerical simulations will be reported in the future.

$Acknowledgments$:-
%\begin{Acknowledgments} 
The authors would like to thank Dr. Tamas Borzsonyi, Prof. Raul Cruz Hidalgo and Mr. Tivadar Pongo for valuable discussions and constructive comments. This research is supported by the Ministry of Science and Technology of the Republic of China grants \#: MOST-107-2112-M-001-025.
%\end{Acknowledgments}

\bibliography{mvexit}

\begin{thebibliography}{23}%
\makeatletter
\providecommand \@ifxundefined [1]{%
 \@ifx{#1\undefined}
}%
\providecommand \@ifnum [1]{%
 \ifnum #1\expandafter \@firstoftwo
 \else \expandafter \@secondoftwo
 \fi
}%
\providecommand \@ifx [1]{%
 \ifx #1\expandafter \@firstoftwo
 \else \expandafter \@secondoftwo
 \fi
}%
\providecommand \natexlab [1]{#1}%
\providecommand \enquote  [1]{``#1''}%
\providecommand \bibnamefont  [1]{#1}%
\providecommand \bibfnamefont [1]{#1}%
\providecommand \citenamefont [1]{#1}%
\providecommand \href@noop [0]{\@secondoftwo}%
\providecommand \href [0]{\begingroup \@sanitize@url \@href}%
\providecommand \@href[1]{\@@startlink{#1}\@@href}%
\providecommand \@@href[1]{\endgroup#1\@@endlink}%
\providecommand \@sanitize@url [0]{\catcode `\\12\catcode `\$12\catcode
  `\&12\catcode `\#12\catcode `\^12\catcode `\_12\catcode `\%12\relax}%
\providecommand \@@startlink[1]{}%
\providecommand \@@endlink[0]{}%
\providecommand \url  [0]{\begingroup\@sanitize@url \@url }%
\providecommand \@url [1]{\endgroup\@href {#1}{\urlprefix }}%
\providecommand \urlprefix  [0]{URL }%
\providecommand \Eprint [0]{\href }%
\providecommand \doibase [0]{http://dx.doi.org/}%
\providecommand \selectlanguage [0]{\@gobble}%
\providecommand \bibinfo  [0]{\@secondoftwo}%
\providecommand \bibfield  [0]{\@secondoftwo}%
\providecommand \translation [1]{[#1]}%
\providecommand \BibitemOpen [0]{}%
\providecommand \bibitemStop [0]{}%
\providecommand \bibitemNoStop [0]{.\EOS\space}%
\providecommand \EOS [0]{\spacefactor3000\relax}%
\providecommand \BibitemShut  [1]{\csname bibitem#1\endcsname}%
\let\auto@bib@innerbib\@empty
%</preamble>
\bibitem [{\citenamefont {Beverloo}\ \emph {et~al.}(1961)\citenamefont
  {Beverloo}, \citenamefont {Leniger},\ and\ \citenamefont {van~de
  Velde}}]{Beverloo61}%
  \BibitemOpen
  \bibfield  {author} {\bibinfo {author} {\bibfnamefont {W.~A.}\ \bibnamefont
  {Beverloo}}, \bibinfo {author} {\bibfnamefont {H.~A.}\ \bibnamefont
  {Leniger}}, \ and\ \bibinfo {author} {\bibfnamefont {J.}~\bibnamefont {van~de
  Velde}},\ }\href {\doibase http://dx.doi.org/10.1016/0009-2509(61)85030-6}
  {\bibfield  {journal} {\bibinfo  {journal} {Chemical Engineering Science}\
  }\textbf {\bibinfo {volume} {15}},\ \bibinfo {pages} {260} (\bibinfo {year}
  {1961})}\BibitemShut {NoStop}%
\bibitem [{\citenamefont {Ashour}\ \emph {et~al.}(2017)\citenamefont {Ashour},
  \citenamefont {Wegner}, \citenamefont {Trittel}, \citenamefont {Borzsonyi},\
  and\ \citenamefont {Stannarius}}]{Ashour17}%
  \BibitemOpen
  \bibfield  {author} {\bibinfo {author} {\bibfnamefont {A.}~\bibnamefont
  {Ashour}}, \bibinfo {author} {\bibfnamefont {S.}~\bibnamefont {Wegner}},
  \bibinfo {author} {\bibfnamefont {T.}~\bibnamefont {Trittel}}, \bibinfo
  {author} {\bibfnamefont {T.}~\bibnamefont {Borzsonyi}}, \ and\ \bibinfo
  {author} {\bibfnamefont {R.}~\bibnamefont {Stannarius}},\ }\href {<Go to
  ISI>://WOS:000395374600009} {\bibfield  {journal} {\bibinfo  {journal} {Soft
  Matter}\ }\textbf {\bibinfo {volume} {13}},\ \bibinfo {pages} {402} (\bibinfo
  {year} {2017})}\BibitemShut {NoStop}%
\bibitem [{\citenamefont {Hong}\ \emph {et~al.}(2017)\citenamefont {Hong},
  \citenamefont {Kohne}, \citenamefont {Morrell}, \citenamefont {Wang},\ and\
  \citenamefont {Weeks}}]{Hong17}%
  \BibitemOpen
  \bibfield  {author} {\bibinfo {author} {\bibfnamefont {X.}~\bibnamefont
  {Hong}}, \bibinfo {author} {\bibfnamefont {M.}~\bibnamefont {Kohne}},
  \bibinfo {author} {\bibfnamefont {M.}~\bibnamefont {Morrell}}, \bibinfo
  {author} {\bibfnamefont {H.~R.}\ \bibnamefont {Wang}}, \ and\ \bibinfo
  {author} {\bibfnamefont {E.~R.}\ \bibnamefont {Weeks}},\ }\href {<Go to
  ISI>://WOS:000417478200003} {\bibfield  {journal} {\bibinfo  {journal}
  {Physical Review E}\ }\textbf {\bibinfo {volume} {96}} (\bibinfo {year}
  {2017})}\BibitemShut {NoStop}%
\bibitem [{\citenamefont {Tang}\ and\ \citenamefont
  {Behringer}(2016)}]{Tang16}%
  \BibitemOpen
  \bibfield  {author} {\bibinfo {author} {\bibfnamefont {J.~Y.}\ \bibnamefont
  {Tang}}\ and\ \bibinfo {author} {\bibfnamefont {R.~P.}\ \bibnamefont
  {Behringer}},\ }\href {<Go to ISI>://WOS:000379522200010} {\bibfield
  {journal} {\bibinfo  {journal} {Epl}\ }\textbf {\bibinfo {volume} {114}}
  (\bibinfo {year} {2016})}\BibitemShut {NoStop}%
\bibitem [{\citenamefont {Nedderman}\ \emph {et~al.}(1982)\citenamefont
  {Nedderman}, \citenamefont {Tüzün}, \citenamefont {Savage},\ and\
  \citenamefont {Houlsby}}]{Nedderman82}%
  \BibitemOpen
  \bibfield  {author} {\bibinfo {author} {\bibfnamefont {R.~M.}\ \bibnamefont
  {Nedderman}}, \bibinfo {author} {\bibfnamefont {U.}~\bibnamefont {Tüzün}},
  \bibinfo {author} {\bibfnamefont {S.~B.}\ \bibnamefont {Savage}}, \ and\
  \bibinfo {author} {\bibfnamefont {G.~T.}\ \bibnamefont {Houlsby}},\ }\href
  {\doibase http://dx.doi.org/10.1016/0009-2509(82)80029-8} {\bibfield
  {journal} {\bibinfo  {journal} {Chemical Engineering Science}\ }\textbf
  {\bibinfo {volume} {37}},\ \bibinfo {pages} {1597} (\bibinfo {year}
  {1982})}\BibitemShut {NoStop}%
\bibitem [{\citenamefont {Diego Lopez-Rodriguez}\ and\ \citenamefont
  {To}(2019)}]{Lopez19}%
  \BibitemOpen
  \bibfield  {author} {\bibinfo {author} {\bibfnamefont {D.~M. A. G. I.~Z.}\
  \bibnamefont {Diego Lopez-Rodriguez}, \bibfnamefont {Diego~Gella}}\ and\
  \bibinfo {author} {\bibfnamefont {K.}~\bibnamefont {To}},\ }\href@noop {}
  {\bibfield  {journal} {\bibinfo  {journal} {submitted}\ } (\bibinfo {year}
  {2019})}\BibitemShut {NoStop}%
\bibitem [{\citenamefont {To}(2002)}]{To02}%
  \BibitemOpen
  \bibfield  {author} {\bibinfo {author} {\bibfnamefont {K.}~\bibnamefont
  {To}},\ }\href@noop {} {\bibfield  {journal} {\bibinfo  {journal} {CHINESE
  JOURNAL OF PHYSICS}\ }\textbf {\bibinfo {volume} {40}},\ \bibinfo {pages} {8}
  (\bibinfo {year} {2002})}\BibitemShut {NoStop}%
\bibitem [{\citenamefont {Corwin}(2008)}]{Corwin08}%
  \BibitemOpen
  \bibfield  {author} {\bibinfo {author} {\bibfnamefont {E.~I.}\ \bibnamefont
  {Corwin}},\ }\href {\doibase 10.1103/PhysRevE.77.031308} {\bibfield
  {journal} {\bibinfo  {journal} {Phys. Rev. E}\ }\textbf {\bibinfo {volume}
  {77}},\ \bibinfo {pages} {031308} (\bibinfo {year} {2008})}\BibitemShut
  {NoStop}%
\bibitem [{\citenamefont {To}\ \emph {et~al.}(2001)\citenamefont {To},
  \citenamefont {Lai},\ and\ \citenamefont {Pak}}]{To01}%
  \BibitemOpen
  \bibfield  {author} {\bibinfo {author} {\bibfnamefont {K.}~\bibnamefont
  {To}}, \bibinfo {author} {\bibfnamefont {P.-Y.}\ \bibnamefont {Lai}}, \ and\
  \bibinfo {author} {\bibfnamefont {H.~K.}\ \bibnamefont {Pak}},\ }\href
  {http://link.aps.org/doi/10.1103/PhysRevLett.86.71} {\bibfield  {journal}
  {\bibinfo  {journal} {Physical Review Letters}\ }\textbf {\bibinfo {volume}
  {86}},\ \bibinfo {pages} {71} (\bibinfo {year} {2001})}\BibitemShut {NoStop}%
\bibitem [{\citenamefont {Zuriguel}\ \emph {et~al.}(2003)\citenamefont
  {Zuriguel}, \citenamefont {Pugnaloni}, \citenamefont {Garcimartin},\ and\
  \citenamefont {Maza}}]{Zuriguel03}%
  \BibitemOpen
  \bibfield  {author} {\bibinfo {author} {\bibfnamefont {I.}~\bibnamefont
  {Zuriguel}}, \bibinfo {author} {\bibfnamefont {L.~A.}\ \bibnamefont
  {Pugnaloni}}, \bibinfo {author} {\bibfnamefont {A.}~\bibnamefont
  {Garcimartin}}, \ and\ \bibinfo {author} {\bibfnamefont {D.}~\bibnamefont
  {Maza}},\ }\href {http://link.aps.org/doi/10.1103/PhysRevE.68.030301}
  {\bibfield  {journal} {\bibinfo  {journal} {Physical Review E}\ }\textbf
  {\bibinfo {volume} {68}},\ \bibinfo {pages} {030301} (\bibinfo {year}
  {2003})}\BibitemShut {NoStop}%
\bibitem [{\citenamefont {Zuriguel}\ \emph {et~al.}(2005)\citenamefont
  {Zuriguel}, \citenamefont {Garcimartin}, \citenamefont {Maza}, \citenamefont
  {Pugnaloni},\ and\ \citenamefont {Pastor}}]{Zuriguel05}%
  \BibitemOpen
  \bibfield  {author} {\bibinfo {author} {\bibfnamefont {I.}~\bibnamefont
  {Zuriguel}}, \bibinfo {author} {\bibfnamefont {A.}~\bibnamefont
  {Garcimartin}}, \bibinfo {author} {\bibfnamefont {D.}~\bibnamefont {Maza}},
  \bibinfo {author} {\bibfnamefont {L.~A.}\ \bibnamefont {Pugnaloni}}, \ and\
  \bibinfo {author} {\bibfnamefont {J.~M.}\ \bibnamefont {Pastor}},\ }\href
  {http://link.aps.org/doi/10.1103/PhysRevE.71.051303} {\bibfield  {journal}
  {\bibinfo  {journal} {Physical Review E}\ }\textbf {\bibinfo {volume} {71}},\
  \bibinfo {pages} {051303} (\bibinfo {year} {2005})}\BibitemShut {NoStop}%
\bibitem [{\citenamefont {To}(2005)}]{To05}%
  \BibitemOpen
  \bibfield  {author} {\bibinfo {author} {\bibfnamefont {K.}~\bibnamefont
  {To}},\ }\href {http://link.aps.org/doi/10.1103/PhysRevE.71.060301}
  {\bibfield  {journal} {\bibinfo  {journal} {Physical Review E}\ }\textbf
  {\bibinfo {volume} {71}},\ \bibinfo {pages} {060301} (\bibinfo {year}
  {2005})}\BibitemShut {NoStop}%
\bibitem [{\citenamefont {Janda}\ \emph {et~al.}(2008)\citenamefont {Janda},
  \citenamefont {Zuriguel}, \citenamefont {Garcimartín}, \citenamefont
  {Pugnaloni},\ and\ \citenamefont {Maza}}]{Janda08}%
  \BibitemOpen
  \bibfield  {author} {\bibinfo {author} {\bibfnamefont {A.}~\bibnamefont
  {Janda}}, \bibinfo {author} {\bibfnamefont {I.}~\bibnamefont {Zuriguel}},
  \bibinfo {author} {\bibfnamefont {A.}~\bibnamefont {Garcimartín}}, \bibinfo
  {author} {\bibfnamefont {L.~A.}\ \bibnamefont {Pugnaloni}}, \ and\ \bibinfo
  {author} {\bibfnamefont {D.}~\bibnamefont {Maza}},\ }\href
  {http://stacks.iop.org/0295-5075/84/i=4/a=44002} {\bibfield  {journal}
  {\bibinfo  {journal} {EPL (Europhysics Letters)}\ }\textbf {\bibinfo {volume}
  {84}},\ \bibinfo {pages} {44002} (\bibinfo {year} {2008})}\BibitemShut
  {NoStop}%
\bibitem [{\citenamefont {Thomas}\ and\ \citenamefont
  {Durian}(2015)}]{Thomas15}%
  \BibitemOpen
  \bibfield  {author} {\bibinfo {author} {\bibfnamefont {C.~C.}\ \bibnamefont
  {Thomas}}\ and\ \bibinfo {author} {\bibfnamefont {D.~J.}\ \bibnamefont
  {Durian}},\ }\href {http://link.aps.org/doi/10.1103/PhysRevLett.114.178001}
  {\bibfield  {journal} {\bibinfo  {journal} {Physical Review Letters}\
  }\textbf {\bibinfo {volume} {114}},\ \bibinfo {pages} {178001} (\bibinfo
  {year} {2015})}\BibitemShut {NoStop}%
\bibitem [{\citenamefont {Merrigan}\ \emph {et~al.}(2018)\citenamefont
  {Merrigan}, \citenamefont {Birwa}, \citenamefont {Tewari},\ and\
  \citenamefont {Chakraborty}}]{Merrigan18}%
  \BibitemOpen
  \bibfield  {author} {\bibinfo {author} {\bibfnamefont {C.}~\bibnamefont
  {Merrigan}}, \bibinfo {author} {\bibfnamefont {S.~K.}\ \bibnamefont {Birwa}},
  \bibinfo {author} {\bibfnamefont {S.}~\bibnamefont {Tewari}}, \ and\ \bibinfo
  {author} {\bibfnamefont {B.}~\bibnamefont {Chakraborty}},\ }\href {\doibase
  10.1103/PhysRevE.97.040901} {\bibfield  {journal} {\bibinfo  {journal}
  {Physical Review E}\ }\textbf {\bibinfo {volume} {97}},\ \bibinfo {pages}
  {040901} (\bibinfo {year} {2018})}\BibitemShut {NoStop}%
\bibitem [{\citenamefont {To}\ and\ \citenamefont {Tai}(2017)}]{To17}%
  \BibitemOpen
  \bibfield  {author} {\bibinfo {author} {\bibfnamefont {K.}~\bibnamefont
  {To}}\ and\ \bibinfo {author} {\bibfnamefont {H.-T.}\ \bibnamefont {Tai}},\
  }\href {https://link.aps.org/doi/10.1103/PhysRevE.96.032906} {\bibfield
  {journal} {\bibinfo  {journal} {Physical Review E}\ }\textbf {\bibinfo
  {volume} {96}},\ \bibinfo {pages} {032906} (\bibinfo {year}
  {2017})}\BibitemShut {NoStop}%
\bibitem [{\citenamefont {Hilton}\ and\ \citenamefont
  {Cleary}(2010)}]{hilton2010effect}%
  \BibitemOpen
  \bibfield  {author} {\bibinfo {author} {\bibfnamefont {J.}~\bibnamefont
  {Hilton}}\ and\ \bibinfo {author} {\bibfnamefont {P.}~\bibnamefont
  {Cleary}},\ }\href@noop {} {\bibfield  {journal} {\bibinfo  {journal}
  {Physics of Fluids}\ }\textbf {\bibinfo {volume} {22}},\ \bibinfo {pages}
  {071701} (\bibinfo {year} {2010})}\BibitemShut {NoStop}%
\bibitem [{\citenamefont {Thomas}\ and\ \citenamefont
  {Durian}(2013)}]{Thomas13}%
  \BibitemOpen
  \bibfield  {author} {\bibinfo {author} {\bibfnamefont {C.~C.}\ \bibnamefont
  {Thomas}}\ and\ \bibinfo {author} {\bibfnamefont {D.~J.}\ \bibnamefont
  {Durian}},\ }\href {http://link.aps.org/doi/10.1103/PhysRevE.87.052201}
  {\bibfield  {journal} {\bibinfo  {journal} {Physical Review E}\ }\textbf
  {\bibinfo {volume} {87}},\ \bibinfo {pages} {052201} (\bibinfo {year}
  {2013})}\BibitemShut {NoStop}%
\bibitem [{\citenamefont {Mankoc}\ \emph {et~al.}(2007)\citenamefont {Mankoc},
  \citenamefont {Janda}, \citenamefont {Arévalo}, \citenamefont {Pastor},
  \citenamefont {Zuriguel}, \citenamefont {Garcimartín},\ and\ \citenamefont
  {Maza}}]{Mankoc07}%
  \BibitemOpen
  \bibfield  {author} {\bibinfo {author} {\bibfnamefont {C.}~\bibnamefont
  {Mankoc}}, \bibinfo {author} {\bibfnamefont {A.}~\bibnamefont {Janda}},
  \bibinfo {author} {\bibfnamefont {R.}~\bibnamefont {Arévalo}}, \bibinfo
  {author} {\bibfnamefont {J.~M.}\ \bibnamefont {Pastor}}, \bibinfo {author}
  {\bibfnamefont {I.}~\bibnamefont {Zuriguel}}, \bibinfo {author}
  {\bibfnamefont {A.}~\bibnamefont {Garcimartín}}, \ and\ \bibinfo {author}
  {\bibfnamefont {D.}~\bibnamefont {Maza}},\ }\href {\doibase
  10.1007/s10035-007-0062-2} {\bibfield  {journal} {\bibinfo  {journal}
  {Granular Matter}\ }\textbf {\bibinfo {volume} {9}},\ \bibinfo {pages} {407}
  (\bibinfo {year} {2007})}\BibitemShut {NoStop}%
\bibitem [{\citenamefont {Mankoc}\ \emph {et~al.}(2009)\citenamefont {Mankoc},
  \citenamefont {Garcimartin}, \citenamefont {Zuriguel}, \citenamefont {Maza},\
  and\ \citenamefont {Pugnaloni}}]{Mankoc09}%
  \BibitemOpen
  \bibfield  {author} {\bibinfo {author} {\bibfnamefont {C.}~\bibnamefont
  {Mankoc}}, \bibinfo {author} {\bibfnamefont {A.}~\bibnamefont {Garcimartin}},
  \bibinfo {author} {\bibfnamefont {I.}~\bibnamefont {Zuriguel}}, \bibinfo
  {author} {\bibfnamefont {D.}~\bibnamefont {Maza}}, \ and\ \bibinfo {author}
  {\bibfnamefont {L.~A.}\ \bibnamefont {Pugnaloni}},\ }\href
  {http://link.aps.org/doi/10.1103/PhysRevE.80.011309} {\bibfield  {journal}
  {\bibinfo  {journal} {Physical Review E}\ }\textbf {\bibinfo {volume} {80}},\
  \bibinfo {pages} {011309} (\bibinfo {year} {2009})}\BibitemShut {NoStop}%
\bibitem [{\citenamefont {Pacheco-V\'azquez}\ \emph {et~al.}(2017)\citenamefont
  {Pacheco-V\'azquez}, \citenamefont {Ramos-Reyes},\ and\ \citenamefont
  {Hidalgo-Caballero}}]{Pacheco17}%
  \BibitemOpen
  \bibfield  {author} {\bibinfo {author} {\bibfnamefont {F.}~\bibnamefont
  {Pacheco-V\'azquez}}, \bibinfo {author} {\bibfnamefont {A.~Y.}\ \bibnamefont
  {Ramos-Reyes}}, \ and\ \bibinfo {author} {\bibfnamefont {S.}~\bibnamefont
  {Hidalgo-Caballero}},\ }\href {\doibase 10.1103/PhysRevE.96.022901}
  {\bibfield  {journal} {\bibinfo  {journal} {Phys. Rev. E}\ }\textbf {\bibinfo
  {volume} {96}},\ \bibinfo {pages} {022901} (\bibinfo {year}
  {2017})}\BibitemShut {NoStop}%
\bibitem [{\citenamefont {Pongo}()}]{Pongo}%
  \BibitemOpen
  \bibfield  {author} {\bibinfo {author} {\bibfnamefont {T.}~\bibnamefont
  {Pongo}},\ }\href@noop {} {}\bibinfo {howpublished} {private
  communication}\BibitemShut {NoStop}%
\bibitem [{\citenamefont {Hidalgo}()}]{Hidalgo}%
  \BibitemOpen
  \bibfield  {author} {\bibinfo {author} {\bibfnamefont {C.~R.}\ \bibnamefont
  {Hidalgo}},\ }\href@noop {} {}\bibinfo {howpublished} {private
  communication}\BibitemShut {NoStop}%
\end{thebibliography}%
\end{document}